\documentclass[letter]{aa}  

\usepackage{graphicx}
\usepackage{txfonts}
\usepackage{hyperref}
\def\jo#1{{{#1}}}
\def\jok#1{{{#1}}}

\begin{document}

   \title{Solar origins of a strong stealth CME detected by Solar Orbiter}
   

   \author{Jennifer O'Kane
          \inst{1}
          \and Lucie M.~Green \inst{1}
          \and Emma E.~Davies \inst{2}
          \and Christian M\"ostl\inst{3}
          \and J\"urgen Hinterreiter \inst{3,4}
          \and Johan L.~Freiherr von Forstner \inst{5}
          \and Andreas J.~Weiss \inst{3,4}
          \and David M.~Long \inst{1}
          \and Tanja Amerstorfer  \inst{3}
          }

   \institute{Mullard Space Science Laboratory, UCL, Holmbury St Mary, Dorking, Surrey, RH5 6NT, UK
              \email{ucasjro@ucl.ac.uk}
              \and Department of Physics, Imperial College London, London, UK
              \and Space Research Institute, Austrian Academy of Sciences, Graz, Austria
              \and Institute of Physics, University of Graz, Graz, Austria
              \and Institut für Experimentelle und Angewandte Physik, Christian-Albrechts-Universität zu Kiel, Kiel, Germany
             }


 
  \abstract
 {}
 {We aim to locate the origin of a stealth coronal mass ejection (CME) detected in situ by the MAG instrument on board Solar Orbiter, and make connections between the CME observed at the Sun, and the \jo{interplanetary CME (ICME)} measured in situ.}
 {Remote sensing data \jo{are} analysed using advanced image processing techniques to identify the source region of the stealth CME, and the global magnetic field at the time of the eruption is examined using \jo{Potential Field Source Surface (PFSS)} models. The observations of the stealth CME at the Sun are compared with the magnetic field measured by the Solar Orbiter spacecraft, and plasma properties measured by the Wind spacecraft.}
 {The source of the CME is found to be a quiet Sun cavity in the northern hemisphere. We find that the stealth CME has a strong magnetic field in situ, despite originating from a quiet Sun region with an extremely weak magnetic field.}
{The interaction of the ICME with its surrounding environment is the likely cause of a higher magnetic field strength measured in situ. Stealth CMEs require multi-wavelength and multi-viewpoint observations in order to confidently locate the source region, however their elusive signatures still pose many problems for space weather forecasting. The findings have implications for Solar Orbiter observing sequences with instruments such as EUI that are designed to capture stealth CMEs.}


   \maketitle
%

\section{Introduction}

Coronal mass ejections (CMEs) are large eruptions of the Sun's plasma and magnetic field into the heliosphere. If Earth-directed, these events have the potential to \jo{drive} space weather effects. A subset of \jo{Interplanetary CMEs (ICMEs)}, known as magnetic clouds, exhibit enhanced magnetic field strength, a rotation in \jo{at least} one of the magnetic field components, and low plasma density, temperature and plasma $\beta$ when detected in situ \citep{burlaga1981}. The helical field configuration of a magnetic cloud is known as a flux rope. The geoeffectiveness of a magnetic cloud (or indeed any CME) is highly dependent on the strength and orientation of the magnetic field. In particular, a strong southward B$_{z}$ for a period of time longer than a few hours typically produces a strong geomagnetic storm. In order to improve space weather prediction capabilities, it is necessary then to study the CME source region and use this knowledge to forecast the configuration and properties of the magnetic field in the interplanetary medium. The source region may or may not itself have a flux rope configuration \cite[see][for reviews]{green2018origin,patsourakos2020}. However, if a flux rope is present it can be identified using a range of remote sensing data and the knowledge used to forecast its geoeffectiveness \citep[see][and references therein]{palmerio2017determining,palmerio2018}.

Stealth CMEs lack the classic low coronal signatures typically associated with CMEs, such as solar flares, \jo{Extreme Ultraviolet (EUV)} dimmings and filament eruptions \citep[see overview by][]{howard2013stealth}. However, it has been shown that image processing techniques can be used to identify observational signatures, although these can be very faint and hard to detect \citep{alzate2017identification,o2019stealth}. Studies carried out so far indicate that stealth CMEs may originate at relatively high heights $\approx$ 1.3~R$_{\odot}$ in the corona \citep{robbrecht2009no,O_Kane_2021}, implying that the eruptive structure has relatively low magnetic field strength and low plasma density, leading to a correspondingly low energy release, lack of strong radiative emissions and lower propagation speed \citep[<500~km/s,][]{d2014observational}. Despite these characteristics, stealth CMEs still have the potential to be geoeffective and therefore present a challenge for space weather forecasting \citep{kilpua2014solar,nitta2017earth}. Carrying out in situ and remote sensing connection studies can help understand the origins, evolution and impact of CMEs in the heliosphere.

Here we study the \jo{solar origins of the first CME detected in situ by \jok{Solar Orbiter \citep{muller2020solar}}. The encounter took place on 19 April 2020 at a heliocentric distance of 0.809~AU}. The details of the CME's structure and propagation in the inner heliosphere are reported in \citet{davies2021} and the \jo{corresponding} Forbush decrease in \citet{Forstner2021}. The interplanetary CME had a clear flux rope structure and its arrival at Earth produced a {\it Dst} minimum of -60~nT on 20 April 2020 and a category G1 geomagnetic storm. At the Sun, the event was determined to be a streamer blowout with no clear signatures in the lower corona. This places the CME in the stealth category and we present here the identification and analysis of the CME source region that was made possible using advanced image processing techniques. The paper is organised as follows: Sect.~\ref{sec:data} describes the in situ and remote sensing data as well as the methods used within this study; Sect.~\ref{sec:results} describes the analysis of the stealth CME connecting the in situ and the solar source perspectives; and finally Sect.~\ref{sec:conclusions} discusses the results and conclusions. 

\section{Data and methods} \label{sec:data}

\subsection{Data}

For the in situ analysis, magnetic field data \jo{were} obtained from the \jok{magnetometer \citep[MAG;][]{horbury2020mag} onboard Solar Orbiter \citep{muller2020solar}}, which provides in situ observations of the CME at approximately 0.8~AU. The Magnetic Field Investigation \citep[MFI;][]{lepping1995} and Solar Wind Experiment \citep[SWE;][]{ogilvie1995} of the Wind spacecraft provide magnetic field and solar wind plasma observations at 1~AU. 

For the remote sensing observations, the white light CME was analysed using data from the Large Angle and Spectrometric Coronagraph \citep[LASCO;][]{brueckner1995large} on board the Solar and Heliospheric Observatory \citep[SOHO;][]{domingo1995soho} and the COR1 and 2 coronagraphs  \citep[part of the Sun Earth Connection Coronal and Heliospheric Investigation instrument suite, SECCHI,][]{howard2008sun} on board the Solar Terrestrial Relations Observatory Ahead \citep[STEREO-A,][]{kaiser2008stereo} spacecraft. EUV data \jo{were} taken from the Atmospheric Imaging Assembly \citep[AIA,][]{lemen2011atmospheric} on board the Solar Dynamics Observatory \citep[SDO,][]{pesnell2011solar} spacecraft, and the Extreme Ultraviolet Imager \citep[EUVI;][]{howard2008sun} onboard STEREO-A. Line of sight magnetic field data \jo{were} taken from the Helioseismic and Magnetic Imager data \citep[HMI,][]{scherrer2012helioseismic} onboard SDO.

\subsection{Methods}
Coronagraph data were processed using the Normalizing-Radial-Graded Filter \citep[NRGF,][]{morgan2006depiction}. EUV data were processed using both the Multi-Scale Gaussian Normalization technique \citep[MGN,][]{morgan2014multi} and the NRGF technique. Stack plots were created using COR1 and COR2 data by taking a radial slice from Sun centre. The slice taken at 90$^\circ$ is presented in this paper, as this most clearly showed the CME curve. The plane-of-sky kinematics were obtained using the Savitzky-Golay bootstrapping technique \citep{byrne2013improved}, and the the 3D shape and orientation of the CME in the COR2 and LASCO/C2 fields of view were modelled using the Graduated Cylindrical Shell model \citep[GCS;][]{thernisien2006modeling,gcs_python}. Longitude and Latitude are defined in Carrington co-ordinates.

\section{The stealth CME} \label{sec:results}

 
 \subsection{In situ observations}
 
  \begin{figure}[t!]
\centering
    \includegraphics[width=1\linewidth]{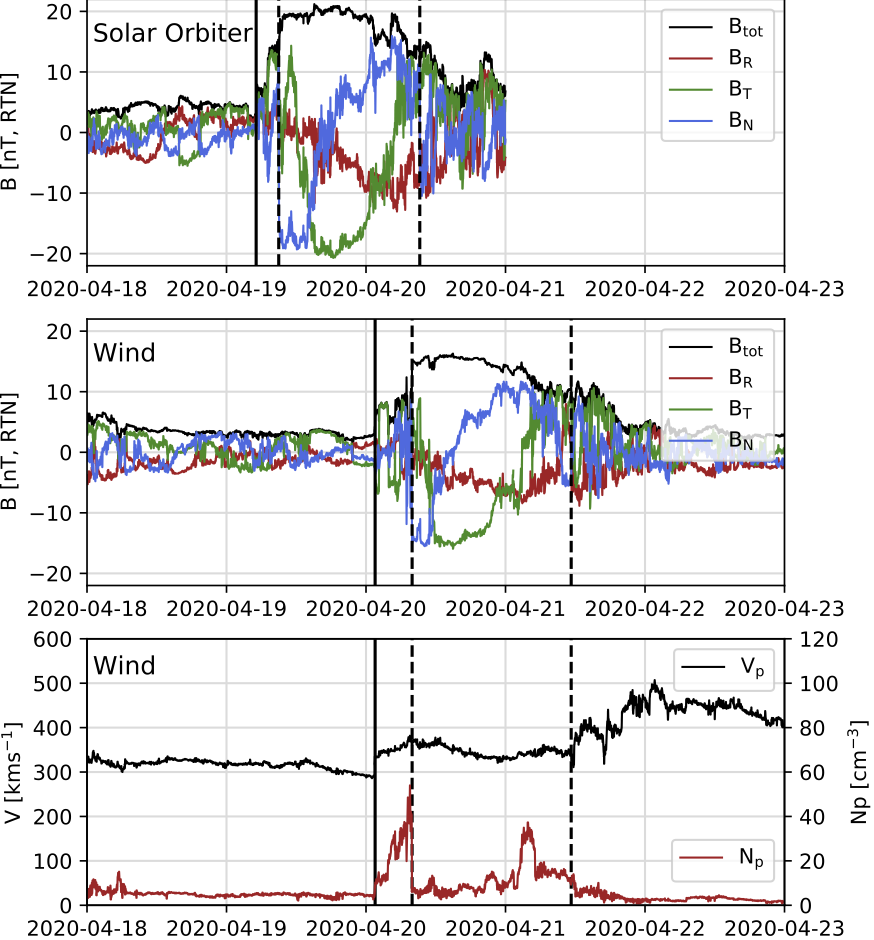}
    \caption{In situ observations of the CME. In each panel, the shock front is delineated by the solid vertical line, and the magnetic flux rope boundaries by the dashed vertical lines. Top panel: Solar Orbiter magnetic field data in Radial-Tangential-Normal (RTN) coordinates, where the total magnetic field magnitude is shown in black and the magnetic field components, RTN, in red, green and blue, respectively. Middle panel: Wind magnetic field data in RTN coordinates, displayed similarly to that of Solar Orbiter. Bottom panel: the proton speed (black) and density (red) measured by Wind.}
    \label{fig:insitu}
  \end{figure}

\begin{figure*}[t!]
\centering
    \includegraphics[width=0.88\linewidth]{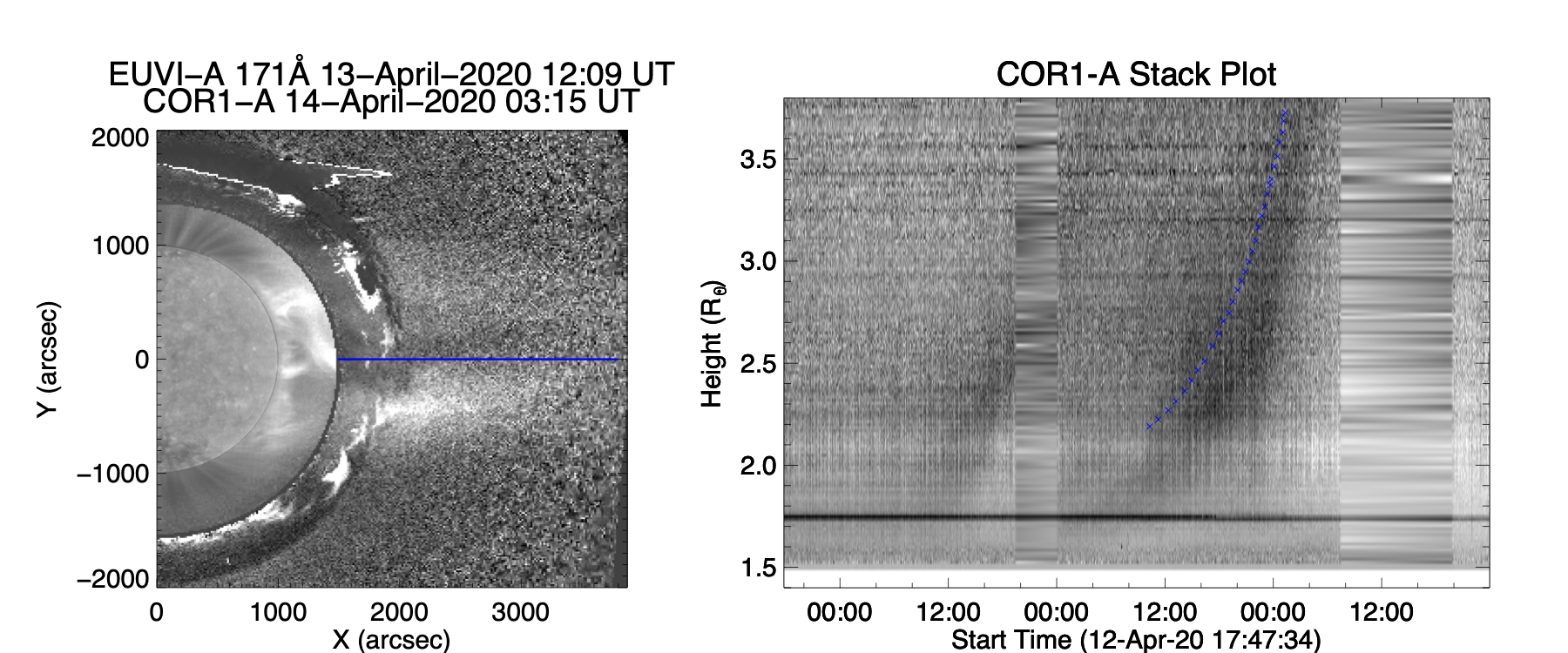}
    \caption{Left: Combined EUVI-A 171\AA\ and COR1-A illustrating the connectivity between the high altitude structures observed in EUVI-A and the CME observed in COR1-A. The EUVI-A image is taken at 22:09~UT 13 April 2020, when the cavity is observed, whilst the COR1-A image is taken at 03:15~UT 14 April 2020, when the CME eruption is underway. The blue line represents the 90$^\circ$ slice taken for the stack plot. Right: A stack plot inverse image of COR1-A, taken at an angle of 90$^\circ$, showing the CME curve. The blue crosses indicate the points selected to track the CME curve.
    }
    \label{fig:stack}
  \end{figure*}
 
In situ magnetic field signatures observed by both Solar Orbiter and Wind, and the solar wind plasma data observed by Wind whilst the spacecraft were close to radial alignment, separated in longitude by less than 5$^{\circ}$ are shown in  Fig. \ref{fig:insitu}. The shock front (delineated by the vertical solid line in Fig. \ref{fig:insitu}) driven by the CME was first observed in situ by Solar Orbiter at 05:06~UT on 19 April 2020 whilst the spacecraft was located at a heliocentric distance of 0.809 AU. The shock was then observed by Wind (0.996 AU) at 01:34~UT on 20 April 2020, arriving with a mean proton speed of 346~km~s$^{-1}$. 
 
The in situ data reveal a clear magnetic flux rope structure, constrained by the vertical dashed lines in Fig. \ref{fig:insitu}, where the leading edge times of the magnetic flux rope are 08:59~UT on 19 April 2020 at Solar Orbiter, and 07:56~UT on 20 April 2020 at Wind \citep[defined by][]{davies2021}. The CME can be classified as a magnetic cloud, meeting criteria detailed by \citet{burlaga1981magnetic}. The configuration of the magnetic flux rope can be classified as South-East-North \citep{bothmer1998,mulligan1998} and therefore has left-handed chirality and the flux rope axis is of low inclination to the ecliptic plane. 


The maximum magnetic field strength reached within the magnetic flux rope was 20.1~nT at Solar Orbiter and 16.3~nT at Wind. This is a relatively high field strength for an ICME at solar minimum, and/or from quiet Sun regions. The proton density of the magnetic flux rope measured at Wind was relatively low throughout, with a mean density of 11.6~cm$^{-3}$, however, this includes the rise in density towards the trailing edge (discounting this rise, the mean density in the flux rope is 8.3~cm$^{-3}$). The rise in density towards the trailing edge is likely due to a compression of the magnetic flux rope caused by the high speed stream that follows, which we observe at Wind to have a mean proton speed of $\sim$450~km~s$^{-1}$. The high speed stream may also be the reason we observe a stronger magnetic field strength than expected within the CME, \jo{where the mean magnetic field strength of the flux rope (18.4~nT at Solar Orbiter and 13.6~nT at Wind) is much stronger than those calculated for ICMEs at similar heliocentric distances \citep[e.g.][]{janvier2019}.}


\jo{The magnetic field profiles observed at both spacecraft show a slight decline in magnetic field strength as the flux rope passes over the spacecraft. Such asymmetric profiles are consistent with the model of an expanding flux rope \citep{nieves2018, janvier2019}.} However, the proton speed profile of the CME observed at Wind shows only a small decline throughout the magnetic flux rope. \jo{Asymmetric magnetic field profiles with a relatively small expansion may correspond to ICMEs with distorted structures \citep{nieves2018}, as observed in images taken by the STEREO-A Heliospheric Imager (HI) of this event \citep{davies2021}.} The mean propagation speed of the leading edge between Solar Orbiter and Wind was 341~km~s$^{-1}$. Assuming a constant transit velocity would suggest a CME onset time $\sim$5 days prior to its arrival at Wind (i.e. on 15 April 2020).

We have applied a semi-empirical 3D flux rope model to the Solar Orbiter data \citep[3DCORE,][]{moestl_2018, weiss2021analysis} in order to derive global parameters of the flux rope within this ICME.  Detailed results are given in \cite{davies2021} and \cite{Forstner2021}, and work on a paper on simultaneous multi-point fitting of the Solar Orbiter, Wind and BepiColombo flux rope magnetic field data is underway (Weiss et al. 2021b, in prep.). Here we present additional information on the magnetic flux derived from the 3DCORE fit on the Solar Orbiter MAG data only. This is of particular interest because for stealth CMEs, large-scale magnetic reconnection during flares can be ruled out as a formation mechanism of the flux rope \citep[e.g.][]{qiu2007, moestl2009}, and yields parameters of a flux rope that was pre-existing in the solar corona.

\cite{davies2021} report the fit results with the 3DCORE technique applied to the Solar Orbiter data, propagated to 1~AU: an axial field strength $B_0=14.3 \pm0.9$~nT, a flux rope diameter of $D_{1AU}=0.114 \pm 0.022$~AU and a twist of $\tau=-3.7 \pm 0.6$ field line turns over the full torus ($\tau <0$ implies left handed chirality), or $\tau_{1AU}= -0.6 \pm 0.1$ turns per AU. We have numerically integrated the flux rope cross section to find an axial flux of $\Phi_{ax}=0.30 \pm 0.06 \times 10^{21}$~Mx. The application of the formula for a uniform twist cylindrical flux rope \citep[e.g. equation 16 in][]{vandas2017}, with the parameters derived from the 3DCORE fit,
\begin{equation}
\label{eq:flux}
\Phi_{ax} = \frac{\pi B_{1AU} }{\tau_{1AU}} \mathrm{ln}(1+\tau_{1AU} D_{1AU}/2)
\end{equation}
gives $\Phi_{ax}=0.33 _{-0.13}^{+0.16} \times 10^{21}$~Mx, which is \jo{consistent} with the numerical result, although this should in general be seen only as a rough approximation because the flux rope cross section \jo{in} 3DCORE is not cylindrical but elliptical. The axial flux we determined for this event is thus at the lower end of reported axial fluxes in ICME flux ropes \citep[cf.][]{qiu2007}.




 \subsection{Stealth CME source region}


 \begin{figure*}[t!]
\centering
    \includegraphics[width=0.88\linewidth]{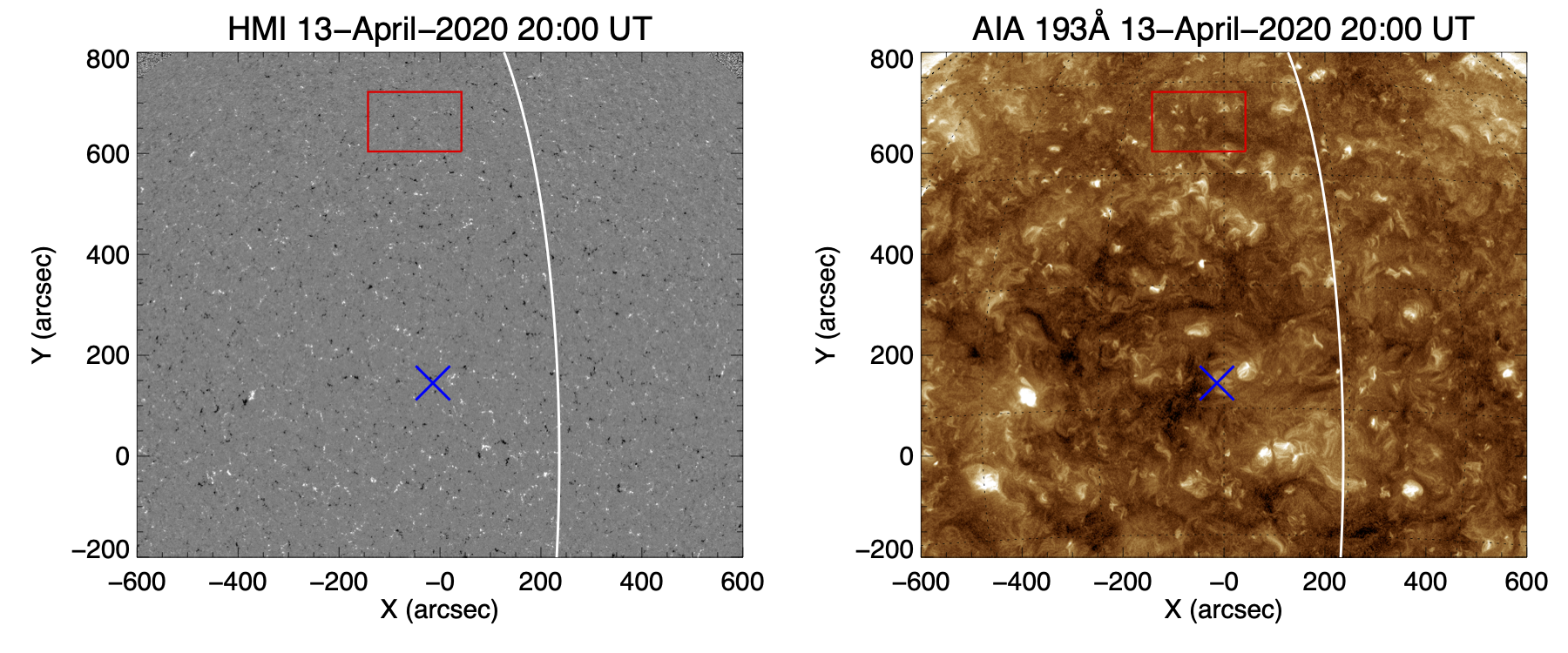}
    \caption{Left: HMI image at 20:00~UT 13 April 2020, showing the magnetic field just before the cavity begins to rise. The blue 'X' marks the location of the GCS approximated source region, whilst the red box marks the approximate location of the cavity that was observed at the solar limb. Right: The corresponding AIA 193\AA\ image, with the GCS and cavity regions highlighted as before. In both figures, the limb of EUVI-A is overplotted as the white line.
    }
    \label{fig:sdo}
  \end{figure*}

\begin{figure}[t!]
\centering
    \includegraphics[width=0.8\linewidth]{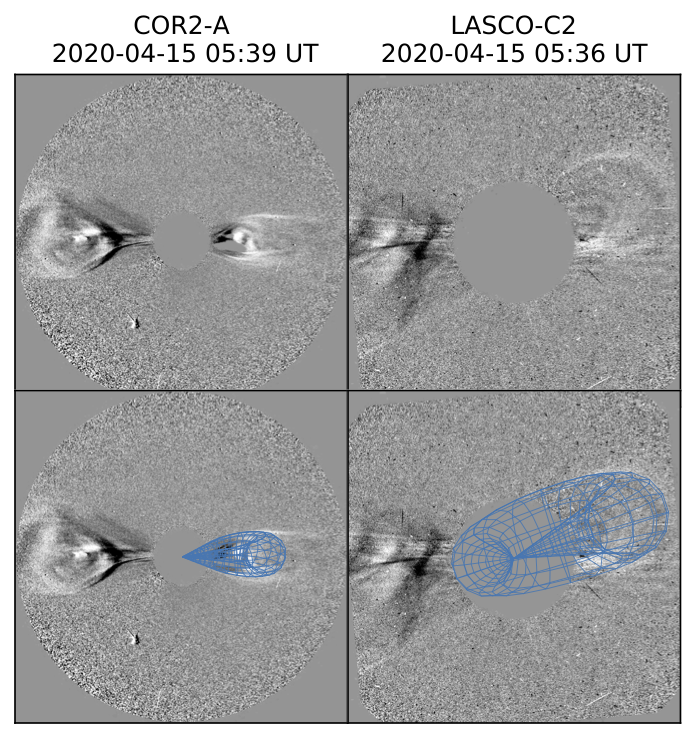}
    \caption{Remote sensing observations of the CME. STEREO-A COR2 (direct images) and SOHO LASCO C2 (running difference images). In the bottom panels, the GCS fitting according to the parameters obtained by \citet{Forstner2021} is indicated by the \jo{blue} mesh.}
    \label{fig:GCS}
  \end{figure}
  
\begin{figure*}[t!]
\centering
    \includegraphics[width=1\linewidth]{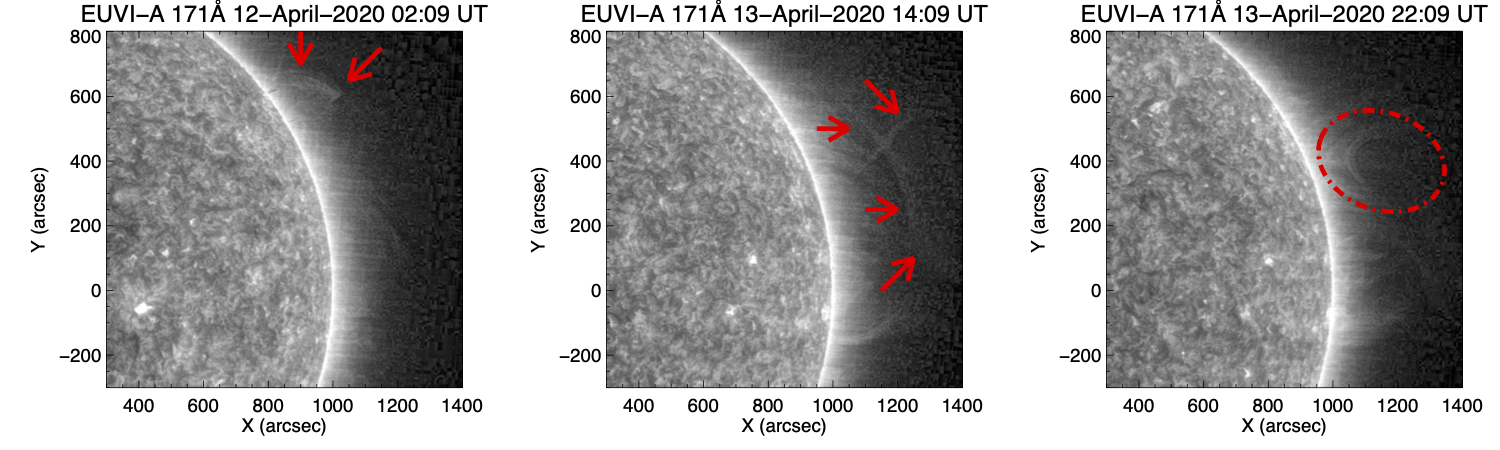}
    \caption{NRGF processed EUVI-A 171\AA\ images, illustrating the structures contributing to the stealth CME. Left panel: The red arrows point out a bright loop structure that is above filament material. Middle panel: The large bright high altitude structure, illustrated by the red arrows. Right panel: The cavity structure, highlighted by the red-dashed circle.
    }
    \label{fig:EUVI}
  \end{figure*}

\begin{figure}[t!]
\centering
    \includegraphics[width=0.88\linewidth]{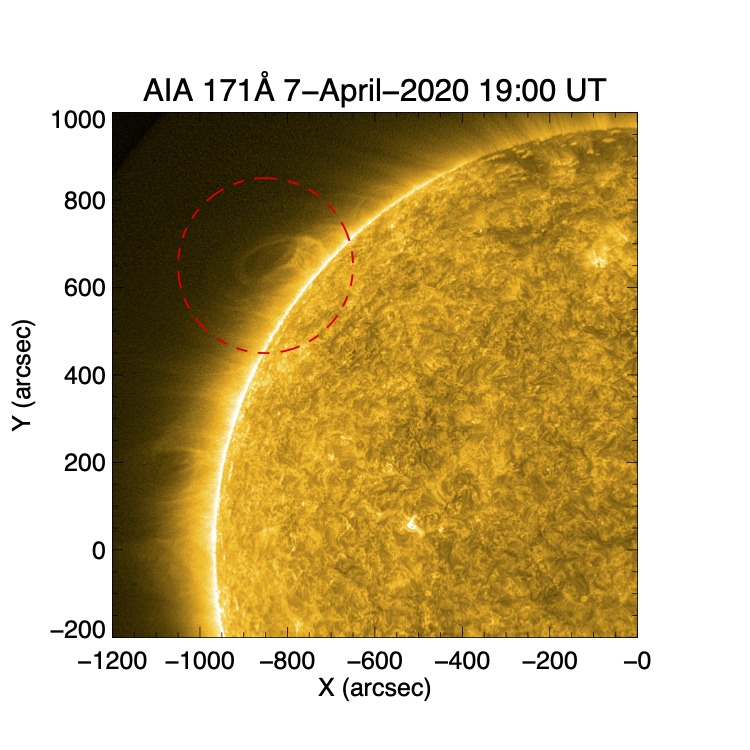}
    \caption{MGN-processed AIA 171\AA\ image at 19:00~UT 7 April 2020, showing the cavity structure observed on the limb (red-dashed circle) $\sim$6 days before the eruption.
    }
    \label{fig:cav}
  \end{figure}

As seen in Fig.\ref{fig:stack}, the stealth CME is observed in COR1-A as a faint enhancement in density, and becomes clear in COR2-A by 03:54~UT 15 April 2020, following the brightening and swelling of a streamer. The CME exhibits a bright underside, and a clear concave-up structure, suggestive of a flux rope configuration \jo{ \citep{vourlidas2013} }. In COR1-A, the underside of the CME can be tracked clearly from 10:16~UT 14 April 2020 (Figure \ref{fig:stack}), using the 90$^\circ$ slice, and has an average velocity of 18~km s$^{-1}$. Using this speed, and a height of 2.5~R$_\odot$, at 22:30~UT 14 April 2020, we estimate a more accurate CME onset time $\sim$19:30~UT 13 April 2020. In COR2-A the CME has an average speed of 53~km s$^{-1}$. \jo{The increase in CME speed between that measured from COR2-A and that measured in situ at Wind, is likely due to the high speed stream following the CME, and the background solar wind.}

Primary observations of the Sun during this time period show no active regions, and the photospheric magnetic field exhibited only weak small-scale field fragments (see Fig.\ \ref{fig:sdo}). EUV and H-alpha data showed that no filaments or filament channels were present either, making the source region of the CME very challenging to identify. Using the images from COR2-A and C2 at 05:39~UT and 05:36~UT 15 April 2020, respectively, the CME was fitted with the GCS model (see Fig.\ref{fig:GCS}). The GCS fit found an approximate source region at 143$\pm$7$^\circ$ longitude, and 3$\pm$3$^\circ$ latitude \citep[see Table 2 in][]{Forstner2021}, and is indicated by the blue cross in Fig.\ \ref{fig:sdo}. This puts the approximate source region on-disk from SDO, and on-limb from STEREO-A.

Although both the MGN and difference imaging techniques were applied to each AIA passband, they did not reveal any clear eruption signatures on disk. However, the NRGF and MGN processed 195~\AA\ and 171~\AA\ EUVI-A data revealed multiple high-altitude dynamic structures off the solar limb; a bright looped region, a large bright structure, and a cavity structure (Fig.\ \ref{fig:EUVI}). The large bright structure is observed from 08:50~UT 13 April 2020, and then slowly rises into COR1-A, coinciding with the brightening and swelling of the streamer. The concave-up cavity structure is observed from 22:09~UT 13 April 2020, at 35.7$^\circ$ latitude, and is deflected southwards as it erupts from the Sun, coinciding with initiation of the CME in COR1-A. We therefore deduce that this is the cavity responsible for the CME that is later detected in situ. The deflection of the cavity was measured by applying the GCS model to the structure in 171~\AA\ EUVI-A. The cavity is observed to deflect southwards by $\sim$19$^\circ$ within the EUVI FOV, and a further $\sim$12$^\circ$ into the FOV of COR1. \jo{It is noted that due to applying the GCS model to one spacecraft instead of two, errors are larger. However, this is smaller for latitude such that the deflection can be measured with reasonable confidence.}

The cavity is an observational signature of a flux rope present in the low corona as the eruption takes place. We track the region back to the \jo{eastern} limb of SDO, to look for evidence of the flux rope present prior to the eruption. Three cavities are observed on the limb in AIA 171~\AA\ between 6 and 10 April 2020. The \jo{first} cavity at 10$^\circ$ latitude erupted off-limb on 8 April 2020, and therefore was ruled out as being the source. The \jok{second cavity at 40$^\circ$ latitude, highlighted by the red-dashed circle in Fig.\ \ref{fig:cav} matches closely with the erupting cavity observed in EUVI-A (at 35.7$^\circ$ latitude) and with its approximate longitude.} \jo{The third cavity at 30$^\circ$ latitude does not match in longitudinal extent with the cavity observed at the limb of EUIV-A. Thus we conclude the second cavity observed is the structure that later erupts as the stealth CME.} Using these two observations, we obtain an approximate source region from the SDO perspective, as 139$^\circ\pm$5$^\circ$ longitude, 38$^\circ\pm$2$^\circ$ latitude, indicated by the red box in Fig.\ \ref{fig:sdo}. The difference between the source region and the estimated region obtained from the GCS is due to the southwards deflection of the cavity as it erupts.

\begin{figure*}[t!]
\centering
    \includegraphics[width=0.88\linewidth,trim=0 100 0 100,clip]{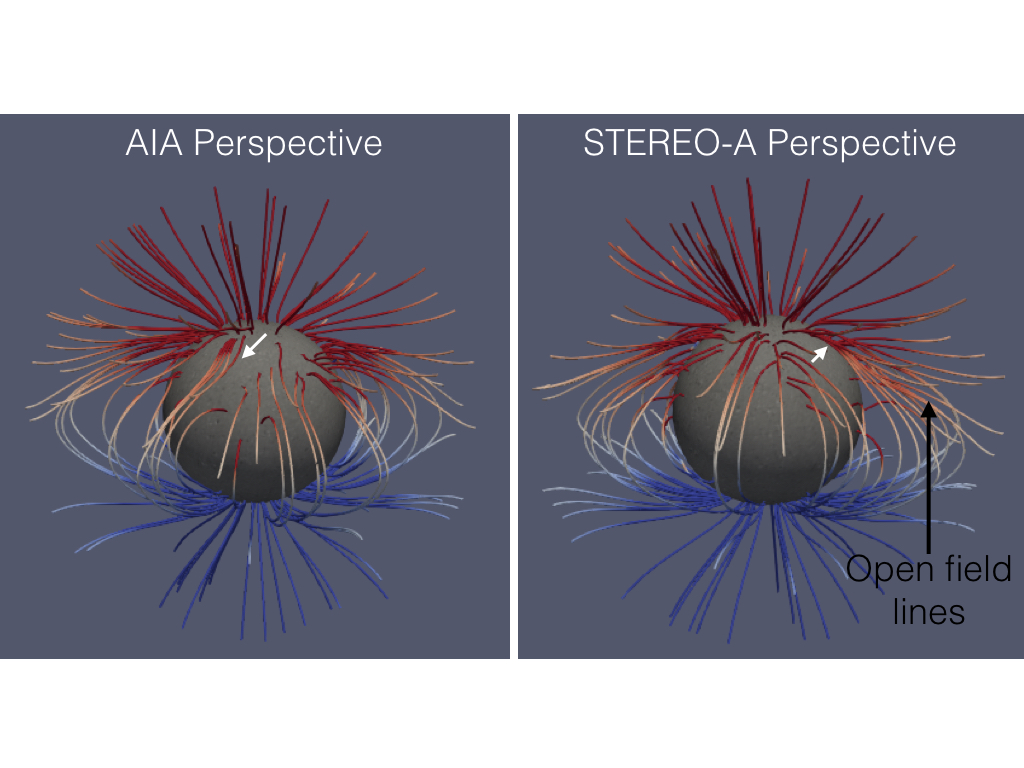}
    \caption{PFSS model of Carrington rotation 2229, as observed from AIA perspective (left) and STEREO-A perspective (right). Positive (negative) polarity is coloured in red (blue), on 13 April 2020. \jo{The approximate source region of the stealth CME is pointed out by the white arrows.}
    }
    \label{fig:pfss}
  \end{figure*}

We further investigated the magnetic topology of the source region and its surrounding environment with a Potential Field Source Surface (PFSS) model, using the PFSS solver in Python \citep[pfsspy,][]{https://doi.org/10.5281/zenodo.1472183,pfss_stansby_2019}. Fig.\ \ref{fig:pfss} shows the PFSS model as viewed from both the AIA (left) and STEREO-A (right) perspectives, with the modelled magnetic field lines showing the positive (red) and negative (blue) polarities. The CME originates in the northern hemisphere, near central meridian (west-limb) the AIA (STEREO-A) perspective. Open field lines are deflected southwards, towards the equator, similar to the trajectory of the cavity observed in EUVI-A. \jo{Given the absence of active regions and solar minimum phase at this time,} the global magnetic field is close to that of a dipole field. 

\section{Discussion and conclusions} \label{sec:conclusions}



We present here a study of the first stealth CME observed in situ by Solar Orbiter, with accompanying Wind in situ observations and STEREO-A and SDO remote sensing data. The combined use of STEREO-A and SDO data revealed that the origin of the CME was a quiet Sun cavity within small-scale magnetic field and no clear polarity inversion line. The cavity identified in EUVI-A and AIA 171~\AA\ passbands extended up to 1.3~R$_{\odot}$ (from Sun centre) in the corona. Such dark elliptical cavities in the quiet Sun indicate the presence of a magnetic flux rope and have been found to have a median height of 1.2~R$_{\odot}$ \citep{gibson2015}. 

Despite the quiet Sun origin of the stealth CME, the magnetic cloud still had a very high field strength in the interplanetary medium (13.6~nT at Wind). The $\sim$1~AU value is significantly higher than that found in a comparison study of stealth CMEs at solar minimum by \citet{kilpua2014solar}, who showed an average B$_{max}$ of 9.7~nT across a sample of 10 stealth CMEs, and the in situ stealth CME study of \cite{nieves2013}. This maximum field strength and the South-East-North configuration of the flux rope may have been the deciding factors in this event producing a category G1 geomagnetic storm. The slow speed of stealth CMEs means we do not typically expect them to drive shocks. 
In this event a forward shock is detected in situ, similar to the case of the stealth CME studied in \citet{nieves2013}. The shock is likely due to the slower solar wind conditions at 1~AU ahead of the ICME ($\sim$300~km~s$^{-1}$). Also, the slight expansion in the unperturbed section of the ICME means that a combination of relatively slow solar wind speed ahead of the CME and the CME expansion may drive the shock as proposed by \citet{lugaz2017}.

This study reinforces the need for multi-wavelength, multi-view point observations of the solar corona, combined with image processing techniques, to further our ability to detect and analyse stealth CME source regions. These observations indicate that for Solar Orbiter EUI to successfully observe stealth CME sources, observing sequences should include regular synoptic images from the 174~\AA\ waveband of the Full Sun Imager (FSI) telescope, which will have a field-of-view of 4 R$_{\odot}$ at perihelion.

\begin{acknowledgements}
       JO thanks the STFC for support via funding given in her PHD studentship. LMG is grateful to the Royal Society, which supported this research through the Royal Society University Research Fellowship scheme. DML is grateful to the Science Technology and Facilities Council for the award of an Ernest Rutherford Fellowship (ST/R003246/1).
       JvF thanks the German Space Agency (Deutsches Zentrum für Luft- und Raumfahrt e.V., DLR) for their support of his work on the Solar Orbiter EPD team under grant 50OT2002. CM, AJW, JH and TA thank the Austrian Science Fund (FWF): P31521-N27, P31659-N27, P31265-N27.
       ED is supported by funding from the Science and Technology Facilities Council (STFC) studentship ST/N504336/1. 
       SDO is a mission of NASA’s Living With a Star Program. STEREO is the third mission in NASA’s Solar Terrestrial Probes program. SOHO is a mission of international cooperation between ESA and NASA. The authors thank the SDO, STEREO, and SOHO teams for making their data publicly accessible.The Solar Orbiter magnetometer was funded by the UK Space Agency (grant ST/T001062/1). Its data are available in the Solar Orbiter Archive at \url{http://soar.esac.esa.int/soar/}.
\end{acknowledgements}

\bibliographystyle{agsm}
\bibliography{Main.bbl}

\end{document}